\documentclass{emulateapj}
\usepackage{amssymb,amsmath,graphicx,longtable,subfigure,wrapfig}
\usepackage{rotating}
\usepackage{tablefootnote}
\usepackage[colorlinks,linkcolor=blue,anchorcolor=green,citecolor=blue]{hyperref}

\begin{document}

\def\func#1{\mathop{\rm #1}\nolimits}
\def\unit#1{\mathord{\thinspace\rm #1}}

\title{Evidence for magnetar formation in broad-lined type Ic supernovae
1998bw and 2002ap}
\author{L. J. Wang\altaffilmark{1}, H. Yu\altaffilmark{2,3}, L. D. Liu%
\altaffilmark{2,3}, S. Q. Wang\altaffilmark{2,3,4}, Y. H. Han\altaffilmark{1}%
, D. Xu\altaffilmark{1}, Z. G. Dai\altaffilmark{2,3}, Y. L. Qiu%
\altaffilmark{1}, J. Y. Wei\altaffilmark{1}}

\begin{abstract}
Broad-lined type Ic supernovae (SNe Ic-BL) are peculiar stellar explosions
that distinguish themselves from ordinary SNe. Some SNe Ic-BL are associated
with long-duration ($\gtrsim 2\unit{s}$) gamma-ray bursts (GRBs). Black
holes and magnetars are two types of compact objects that are hypothesized
to be central engines of GRBs. In spite of decades of investigations, no
direct evidence for the formation of black holes or magnetars has been found
for GRBs so far. Here we report the finding that the early peak ($t\lesssim
50\unit{days}$) and late-time ($t\gtrsim 300\unit{days}$) slow decay
displayed in the light curves of both SNe 1998bw (associated with GRB
980425) and 2002ap (not GRB-associated) can be attributed to magnetar
spin-down with initial rotation period $P_{0}\sim 20\unit{ms}$, while the
intermediate-time ($50\lesssim t\lesssim 300\unit{days}$) exponential
decline is caused by radioactive decay of $^{56}$Ni. The connection between
the early peak and late-time slow decline in the light curves is unexpected
in alternative models. We thus suggest that GRB 980425 and SN 2002ap were
powered by magnetars.
\end{abstract}

\keywords{stars: neutron --- supernovae: general --- supernovae: individual
(SNe 1998bw, 2002ap)}

\affil{\altaffilmark{1}Key Laboratory of Space Astronomy and Technology,
National Astronomical Observatories,
Chinese Academy of Sciences, Beijing 100012, China; wanglj@nao.cas.cn, wjy@nao.cas.cn}

\affil{\altaffilmark{2}School of Astronomy and Space Science, Nanjing
University, Nanjing 210093, China; dzg@nju.edu.cn}

\affil{\altaffilmark{3}Key Laboratory of
Modern Astronomy and Astrophysics (Nanjing University),
Ministry of Education, Nanjing 210093, China}

\affil{\altaffilmark{4}Department of Astronomy, University of  California,
Berkeley, CA 94720-3411, USA}

\section{Introduction}

Broad-lined type Ic supernovae (SNe Ic-BL) are a subclass of core-collapse
type Ic SNe (CCSNe) having broad absorption troughs in their optical
spectra. Since their discovery, SNe Ic-BL stand out as a subclass that is
peculiar compared with other ordinary CCSNe. One of their peculiarities is
that their astonishing kinetic energy ($\sim 10^{52}\unit{erg}$) is
unexpected in the well-studied SN explosion mechanisms %
\citep{Bethe90,Janka16}.

Another peculiarity of SNe Ic-BL is the failure of one-dimensional $^{56}$Ni
model -- which is assumed to work well for most of ordinary SNe Ibc\footnote{%
Currently the interest in ordinary SNe Ibc is low \citep{Clocchiatti11}.
Many SNe Ibc were not studied after observed for a duration, e.g., SNe
2004dn ($90\unit{days}$), 2004fe ($50\unit{days}$), 2004ge ($70\unit{days}$%
), 2004gt ($170\unit{days}$), 2005az ($120\unit{days}$), 2005kl ($160\unit{%
days}$), 2005mf ($110\unit{days}$), 2007cl ($70\unit{days}$), to mention a
few (see https://sne.space/). The bolometric light curves of some
well-observed SNe Ibc, e.g., SNe 1983V \citep{Clocchiatti97b}, 1990B %
\citep{Clocchiatti01}, 1992ar \citep{Clocchiatti00}, are not available. For
other well-observed SNe Ibc, e.g., SNe 1994I \citep{Clocchiatti08}, 2004aw %
\citep{Taubenberger06}, 2007gr \citep{Hunter09,Chen14}, 2011bm %
\citep{Valenti12}, their bolometric light curves do have been constructed
but the light curves were not modeled in detail. For virtually all ordinary
SNe Ibc, it is usually assumed that they were powered by $^{56}$Ni %
\citep{Drout11}.} -- in reproducing their light curves. What is intricate in
reproducing the light curves of SNe Ic-BL is that the simple analytical
models have difficulty in modeling both the peak part and the later part.
Most of the time different $^{56}$Ni masses are derived, or multiple-zones
need to be invoked \citep{Maeda03,Wheeler15}.\footnote{%
The physics behind this is still not clear. Nickel-mixing, asymmetries and
complete hydromodeling with realistic density structures and opacities are
likely needed to resolve the issues. This highlights the caveat in deriving
the $^{56}$Ni mass from the peak $R$-band magnitude suggested by \cite%
{Drout11}.} In light of this difficulty, it was found that the light curve
of SNe Ic-BL can be reasonably reproduced by proposing that the ejecta
consist of two components, i.e. one fast-moving outer component which is
responsible for the early peak and the inner slow compact component which
produces the late-time exponential decline \citep{Maeda03}. Such a
two-component $^{56}$Ni model is partially supported because of the
asphericity suggested by polarization measurements \citep{Patat01} and
peculiar nebular line profiles \citep{Mazzali01,Mazzali05,Maeda02,Maeda08}
of some SNe Ic-BL. Hydrodynamic models of jetlike explosions %
\citep{Nagataki97,Khokhlov99,MacFadyen99,Maeda02} suggest a prolate
spheroid. In such a scenario, a fast-moving component is produced along the
jet, while the compact component is moving slowly along the perpendicular
direction. Because of the energies of these two components are different,
their synthesized $^{56}$Ni masses are usually different.\footnote{%
There are caveats in interpreting the polarization measurements and peculiar
nebular line profiles as asphericity. The measured polarization value $0.5\%$
of SN 1998bw \citep{Patat01} is similar to those for other ordinary CCSNe %
\citep{Wang96}. It was also found that the double-peaked oxygen lines in
nebular spectra of some SNe Ic-BL are common within other types of SNe, e.g.
SNe IIb, Ib, and Ic \citep{Modjaz08}. In addition, some SNe Ic-BL are not
accompanied by powerful relativistic jets 
\citep[e.g., SN
2002ap,][]{Berger02}.}

SNe Ic-BL are also of great astrophysical importance because they are the
only SNe in association with long-duration gamma-ray bursts %
\citep[GRBs;][]{Galama98,Bloom99,Hjorth03,Stanek03,Campana06,Mazzali06,Woosley06,Cano16}%
. Despite persistent investigation, there is still no direct evidence for
the two types of hypothesized central engine of GRBs, i.e. black holes %
\citep{Popham99,Narayan01,Kohri02,Liu07,Liu16,Song16} and/or magnetars %
\citep{Usov92,Dai98a,Dai98b,ZhangD08,ZhangD09,ZhangD10,Mosta15}. GRBs are
usually collimated relativistic phenomena, while SNe Ic-BL are mostly
subrelativistic and nearly isotropic.\footnote{%
There is no evidence for strong collimation in the class of low-luminosity
GRBs \citep[see e.g.,][]{Soderberg06b}. Some SNe Ic-BL show evidence for
mildly relativistic ejecta (SN 2009bb, \citealt{Soderberg10}; SN 2012ap, %
\citealt{Margutti14}, \citealt{Chakraborti15}).} The remnants of all CCSNe
are presumably black holes or neutron stars. The light curve of an SN
powered by a black hole \citep{Dexter13,Gao16} is different from that
powered by a magnetar. Therefore SNe Ic-BL could shed light on the still
elusive central engine of GRBs.

There is indirect evidence that SNe Ic-BL are powered by millisecond
magnetars because their kinetic energy has an upper limit that is close to
the rotational energy of a neutron star spinning at nearly broken-up
frequency \citep{Mazzali14}. Furthermore, the progenitor mass\footnote{%
Theoretical models \citep[see e.g.,][]{Heger03} indicate that the core
collapses of main sequence stars with different initial masses and
metallicity could result in black holes or neutron stars. The low explosion
energy and less mass ejected by SN 2006aj are consistent with a low mass $%
\sim 20M_{\odot }$ main sequence star, which is predicted to result in a
neutron star after core collapse. However, the evolution of a massive star
may be influenced by other factors, e.g. rotation, magnetic field, binary
interaction \citep{Yoon15}, which are not well understood right now.} of SN
2006aj associated with GRB 060218 \citep{Mazzali06} and the light curve of
SN 2011kl associated with GRB 111209A \citep{Greiner15} are consistent with
magnetar formation. Nonetheless, the modeling uncertainty of stellar
evolution for SN 2006aj \citep{Mazzali06} and the short-duration data
coverage ($t\lesssim 60\unit{days}$) and moderate data accuracy for SN
2011kl \citep{Greiner15} make such evidence inconclusive.

In developing a magnetar model for SNe Ic-BL, it is found for SNe 1997ef and
2007ru that the early rapid rise and decline ($t\lesssim 50\unit{days}$) in
the light curves stems from the contribution of a rapidly spinning magnetar,
while the later exponential decline ($50\lesssim t\lesssim 200\unit{days}$)
can be attributed to the $^{56}$Co radioactive decay \citep{Wang16b}. This
model reduces the total $^{56}$Ni mass needed to power the light curve and
therefore solves the long-lasting problem for the magnetar model that the
shock caused by the spin-down of a magnetar cannot synthesize the needed $%
^{56}$Ni \citep{Nishimura15,Suwa15}. In addition, this model naturally
provides the huge kinetic energy of SNe Ic-BL by converting most of the
magnetar's rotational energy into kinetic energy of the ejecta. The
successful demonstration of the magnetar model in reproducing the light
curves of SNe Ic-BL is a supportive evidence for magnetar formation in SNe
Ic-BL. However, we have to bear in mind that the magnetar model is only one
possible choice for SNe Ic-BL because the two-component model is also able
to reproduce reasonably well the light curves of most SNe Ic-BL %
\citep{Maeda03}.

The motivation for this work is, on the one hand, to assess the ability of
the magnetar model in reproducing the very late-phase ($t\gtrsim 300\unit{%
days}$) light curves of SNe Ic-BL when the luminosity decline rate shows
evidence for deviation from the exponential decline (during the phase $%
50\lesssim t\lesssim 200\unit{days}$) and on the other hand to compare it
with the two-component model.

The reason for observational data for $t\gtrsim 300\unit{days}$ is as
follows. If one focuses on the data within one year after explosion, the $%
t^{-2}$ tail in the magnetar model easily parallels that expected from $%
^{56} $Co decay and one cannot unambiguously tell if it is the magnetar or $%
^{56}$Co that is powering the light curve \citep{Woosley10,Inserra13}. Only
at very late phases can one distinguish between magnetar model and $^{56}$Co
decay model.

To accurately determine the parameter values, a Markov chain Monte Carlo
program is developed. We search the literature and find that SNe 1998bw and
2002ap have an observational coverage well beyond $500\unit{days}$ and are
therefore quite suitable for our investigation. To our surprise, in Section %
\ref{sec:data} it is found that in the magnetar model, the magnetar not only
contributes to the early peak of the light curve of the broad-lined SNe
1998bw and 2002ap, but it can also manifest itself as a significant excess
over the exponential decay in late-time light curves when the $^{56}$Co
contribution became small compared to the magnetar spin-down luminosity. The
results are discussed in Section \ref{sec:dis} and it is argued that SNe
1998bw and 2002ap provide hitherto strong evidence for magnetar formation in
SNe Ic-BL. A short summary is given in Section \ref{sec:con}.

\section{Data preparation and modeling}

\label{sec:data}

Being one of the nearest SNe in the last decades, SN 2002ap triggered an
observational campaign since its discovery on 2002 January 29 %
\citep{Gal-Yam02,Mazzali02,Foley03,Yoshii03,Tomita06}. Thanks to its
proximity, only $9.4\unit{Mpc}$ in distance, high quality observational data
were acquired until $580\unit{days}$ after explosion, which is prerequisite
for the identification of late excess over the exponential $^{56}$Co decay.

SN 1998bw \citep{Galama98,McKenzie99,Sollerman00,Patat01,Clocchiatti11}, on
the other hand, is the nearest SN associated with a GRB \citep{Cano16}.
Although at a distance greater than SN 2002ap, its brighter luminosity
qualifies SN 1998bw as an ideal observational target and its light curve was
measured to $\sim 1000\unit{days}$ post explosion \citep{Sollerman02}.

To accurately model the light curve of SNe 1998bw and 2002ap, we develop a
Markov chain Monte Carlo approach based on the recently updated analytic
magnetar model \citep{Wang16c}. In this model, photospheric recession is
considered so that the photospheric velocity evolution is traced. The
acceleration of the SN ejecta by the magnetar energy injection is taken into
account. That is, in this model the kinetic energy of the SNe Ic-BL is
believed to originate mainly from the magnetar spin-down. Nevertheless, this
does not preclude the possibility of a non-zero initial explosion energy.
This model also incorporates the high energy photon leakage (%
\citealt{WangWang15}, see also \citealt{Chen15}) based on the fact that the
energy injection emanated from the magnetar could be dominated by high
energy photons \citep{Buhler14,Wang16a}.

In summary, aside from the usual parameters, e.g. the ejecta mass $M_{%
\mathrm{ej}}$, the $^{56}$Ni mass $M_{\mathrm{Ni}}$, the opacity in optical
band $\kappa $, we also need the opacity $\kappa _{\gamma ,\mathrm{Ni}}$ to $%
^{56}$Ni (including $^{56}$Co) decay photons, the opacity $\kappa _{\gamma ,%
\mathrm{mag}}$ to magnetar photons and the magnetar parameters, i.e. the
dipole magnetic field $B_{p}$ and initial rotation period $P_{0}$.

The optical opacity $\kappa $ is strongly degenerated with the ejecta mass $%
M_{\mathrm{ej}}$ and cannot be accurately determined in this model.
Fortunately, $\kappa $ is a parameter that characterizes the microphysics of
the ejecta and therefore can be calculated in first principles based on our
knowledge about the ejecta composition. It is believed that the progenitor
of an SN Ic (broad-lined or not) is a massive single star or a low-mass star
in a binary \citep{Smartt09}. The ejecta of such a star explosion are mainly
composed of $^{16}$O, $^{20}$Ne and $^{24}$Mg %
\citep{Iwamoto00,Nakamura01a,Maeda02}. In the literature a range of values,
from $0.06\unit{cm}^{2}\unit{g}^{-1}$ to $0.2\unit{cm}^{2}\unit{g}^{-1}$, is
used for such a composition 
\citep[e.g.,][and references
therein]{WangWang15b,Dai16}. Here we adopt the constant value $\kappa =0.1%
\unit{cm}^{2}\unit{g}^{-1}$. We note that the optical opacity is also a
function of the ionization state of the ejecta. A magnetar sitting in the
center of the explosion provides a source of ionizing radiation %
\citep{Metzger14,Wang16a} and therefore the ejecta could be ionized even
when they are cooled to a low temperature. This means that the late-time
ionization level of the ejecta powered by a magnetar is usually higher than
that powered by radioactive decay. A high level of ionization inclines to
keep the optical opacity constant over a long time. We therefore suggest
that a constant optical opacity is appropriate in the magnetar model.

The opacity to $^{56}$Ni decay photons $\kappa _{\gamma ,\mathrm{Ni}}$
usually takes the fiducial value $\kappa _{\gamma ,\mathrm{Ni}}=0.025-0.027%
\unit{cm}^{2}\unit{g}^{-1}$ \citep{Colgate80,Swartz95}. In the actual
applications it is found from case to case that this value cannot fit the
light curve of some SNe %
\citep[e.g.,][]{Filippenko86,Tsvetkov86,Schlegel89,Swartz91,Wang16b}. This
discrepancy may be caused by some macrophysical ignorance, e.g. a peculiar
density distribution of the ejecta. In this work we leave $\kappa _{\gamma ,%
\mathrm{Ni}}$ as a free parameter and appreciate its deviation from the
fiducial value as macrophysical uncertainties.

Another more uncertain parameter is the opacity to magnetar photons $\kappa
_{\gamma ,\mathrm{mag}}$. Depending on the energy spectra of the spinning
magnetar, $\kappa _{\gamma ,\mathrm{mag}}$ may vary from $10^{-2}\unit{cm}%
^{2}\unit{g}^{-1}$ to $10^{6}\unit{cm}^{2}\unit{g}^{-1}$ \citep{Kotera13}.
Based on this fact, we set $\kappa _{\gamma ,\mathrm{mag}}$ as a free
parameter.

In the previous magnetar modeling to demonstrate that the early light curve
of SNe Ic-BL is due to the contribution of a magnetar, we assume that the
kinetic energy of SNe Ic-BL comes exclusively from the magnetar spin-down %
\citep{Wang16b}. This is approximately true for the SNe studied by \cite%
{Wang16b} because the rapid rise in the light curve of SNe 1997ef and 2007ru
precludes a significant initial explosion energy. In a more general case,
however, any SN explosion should have an initial explosion energy.
Consequently, we parameterize the initial explosion energy by including an
initial expansion velocity $v_{\mathrm{sc}0}$ in our model.

\begin{figure*}[tbph]
\centering\includegraphics[width=0.68\textwidth,angle=0]{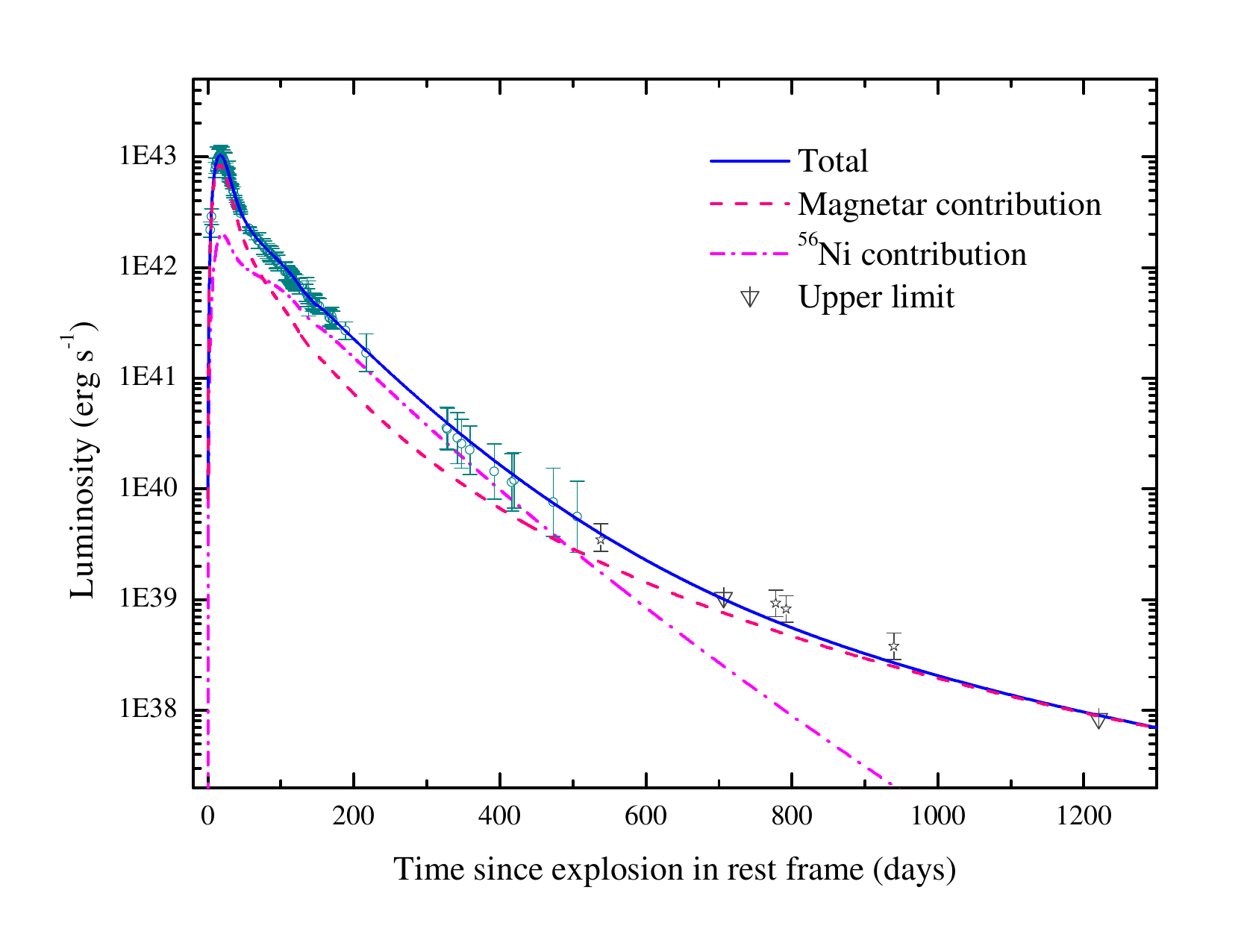}
\caption{Best-fit light curve (solid line) of SN 1998bw. The dashed line
(the magnetar contribution) is the light curve with $M_{\mathrm{Ni}}=0$
while other parameters take the same values as the solid line. The
dot-dashed line is the difference between these two light curves, i.e., the $%
^{56}$Ni contribution. The data points ($t<506\unit{days}$) are taken from 
\protect\cite{Patat01}. In this figure we do not try to fit the dark star
data points and the two upper limits. They are taken from \protect\cite%
{Sollerman02} after shifted upward by $0.6\unit{dex}$ and plotted here just
for eye guidance. See the text for more explanation.}
\label{fig:1998bw}
\end{figure*}

\begin{figure*}[tbph]
\centering\includegraphics[width=0.68\textwidth,angle=0]{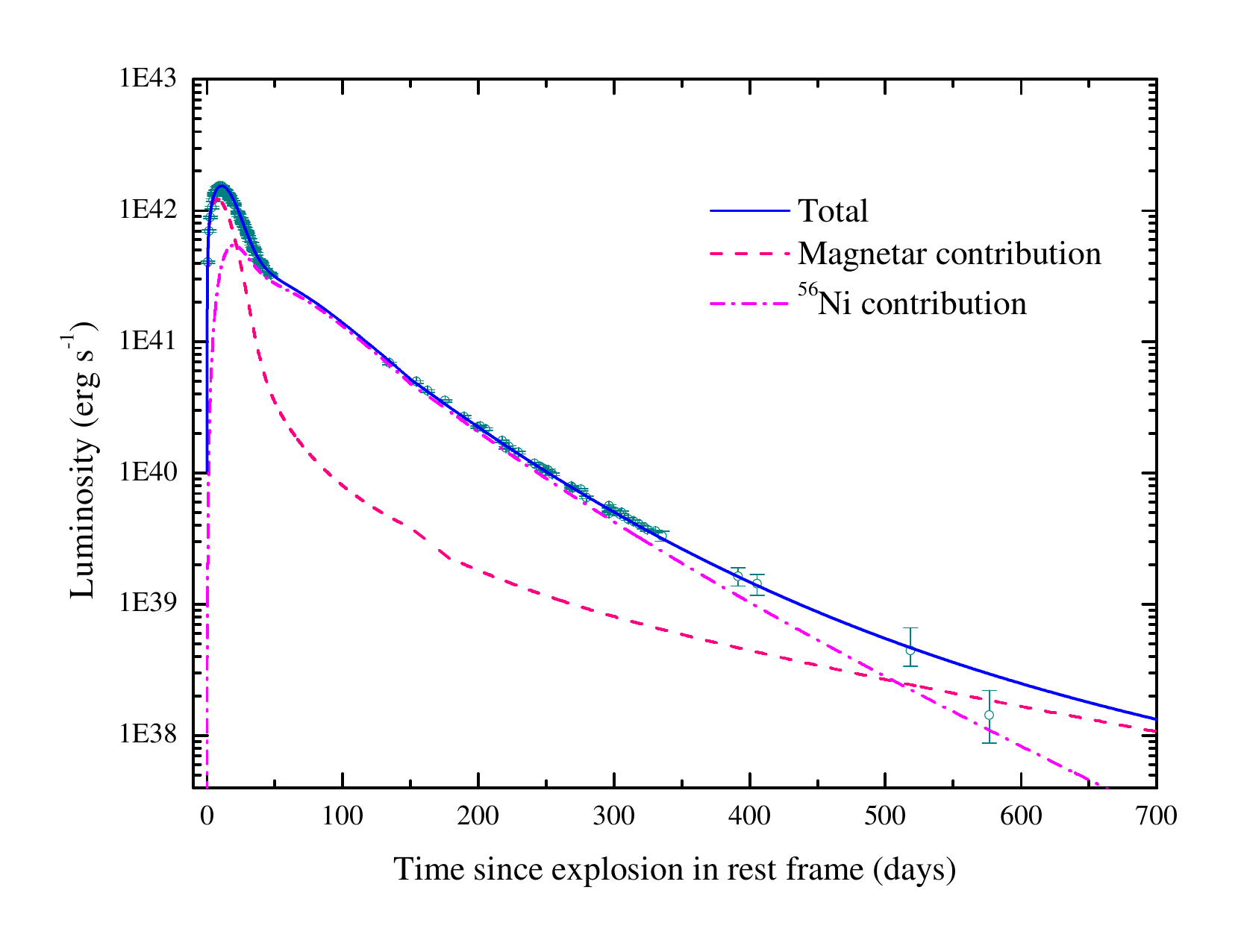}
\caption{Same as Figure \protect\ref{fig:1998bw}, but for SN 2002ap.}
\label{fig:2002ap}
\end{figure*}

\begin{table*}[tbph]
\caption{Best-fit parameters for SNe 1998bw and 2002ap. The uncertainties
are $1\protect\sigma $ errors.}
\label{tbl:para}
\begin{center}
\begin{tabular}{ccccccccc|c}
\hline\hline
SN & $M_{\mathrm{ej}}$\tablenotemark{a} & $M_{\mathrm{Ni}}$ \tablenotemark{b}
& $B_{p}$\tablenotemark{c} & $P_{0}$\tablenotemark{d} & $v_{\mathrm{sc}0}$%
\tablenotemark{e} & $\kappa _{\gamma ,\mathrm{Ni}}$\tablenotemark{f} & $%
\kappa _{\gamma ,\mathrm{mag}}$\tablenotemark{g} & $T_{\mathrm{start}}$%
\tablenotemark{h} & $E_{K}$\tablenotemark{i} \\ 
& $\left( M_{\odot }\right) $ & $\left( M_{\odot }\right) $ & $\left( 10^{15}%
\unit{G}\right) $ & $\left( \unit{ms}\right) $ & $\left( \unit{km}\unit{s}%
^{-1}\right) $ & $\left( \unit{cm}^{2}\unit{g}^{-1}\right) $ & $\left( \unit{%
cm}^{2}\unit{g}^{-1}\right) $ & $\left( \unit{days}\right) $ & $\left(
10^{51}\unit{erg}\right) $ \\ \hline
1998bw & $2.6_{-0.4}^{+0.5}$ & $0.10_{-0.02}^{+0.03}$ & $%
1.66_{-0.14}^{+0.21} $ & $20.8\pm 0.4$ & $11040_{-1591}^{+1501}$ & $%
0.31_{-0.11}^{+0.14}$ & $0.29_{-0.17}^{+0.19}$ & $-0.009_{-0.36}^{+0.32}$ & $%
1.86$ \\ 
2002ap & \multicolumn{1}{l}{$2.0_{-0.36}^{+0.14}$} & $0.029\pm 0.0004$ & $%
13.7\pm 0.3$ & $22.2_{-1.2}^{+1.3}$ & $8134_{-1878}^{+344}$ & $%
0.08_{-0.017}^{+0.003}$ & $5.3_{-2.3}^{+2.0}$ & $1.5_{-0.07}^{+0.06}$ & $%
0.79 $ \\ \hline
\end{tabular}%
\end{center}
\par
\textbf{Note.} In these fits, we fix $\kappa =0.1\unit{cm}^{2}\unit{g}^{-1}$%
. The parameters on the left side of the vertical line are fitting values,
while the one on the right side is derived from the fitting values.
\par
a. Ejecta mass.
\par
b. $^{56}$Ni mass.
\par
c. Magnetic dipole field strength of the magnetar.
\par
d. Initial rotation period of the magnetar.
\par
e. Initial expansion velocity of the ejecta.
\par
f. Opacity to $^{56}$Ni decay photons.
\par
g. Opacity to magnetar spin-down photons.
\par
h. Explosion time relative to the observational data.
\par
i. Initial explosion energy.
\end{table*}

Because we usually do not know the explosion time of the SN, we include a
parameter, the explosion time $T_{\mathrm{start}}$, in the Monte Carlo
program. For SN 1998bw, the explosion time is known and in this case we can
actually evaluate the validity of the fitting procedure by comparing the
determined $T_{\mathrm{start}}$ with the actual explosion time.

The luminosity data points for SN 2002ap are taken from \cite{Tomita06}. The
luminosity observational errors are all taken from Figure 4 of \cite%
{Tomita06}, i.e., $\pm 0.03\unit{mag}$ except for the last five points. The
photospheric velocity data are taken from \cite{Gal-Yam02} and \cite%
{Mazzali02}. The uncertainties in the measurement of photospheric velocity
is usually large \citep[cf. Figure 8 of][]{Valenti08}. In view of this fact,
we adopt the measurement error in photospheric velocity as half of the
measured value.\footnote{%
After the submission of this paper, we have noted that \cite{Modjaz16} have
compiled a large sample of spectra of SNe Ic, broad-lined or not. In this
sample the photospheric velocities of the SNe have been measured in an
improved way. We will assess the difference made by this improvement in
future work.}

The data points for SN 1998bw up to $506\unit{days}$ after explosion,
including the UVOIR bolometric luminosity data and photospheric velocity
data, are taken from \cite{Patat01}.\footnote{\cite{Clocchiatti11} presented
an update on the light curve by providing additional data spanning from $%
\sim 40$ up to $\sim 60\unit{days}$ after the explosion. The fact that the
data points during this period were well sampled by \cite{Patat01} on the
one hand, and \cite{Clocchiatti11} only provided UBV(RI)$_{C}$ band
observations up to $85\unit{days}$, on the other, makes us decide to choose
the data of \cite{Patat01} in this work.} \cite{Sollerman02} obtained OIR
luminosity up to $\sim 1000\unit{days}$ post explosion (their Figure 4).
Because of the missing of $UV$ band data in the light curve of Figure 4 in 
\cite{Sollerman02}, we do not try to fit the data from \cite{Sollerman02}.
Nevertheless, we try to compare the data from \cite{Patat01} and \cite%
{Sollerman02} and find that the data at $t\sim 500\unit{days}$ can be bring
consistent with each other if the data given by \cite{Sollerman02} are
shifted $0.6\unit{dex}$ upward.\footnote{%
It is a common practice to shift a constant factor of individual band
magnitude to obtain bolometric magnitude \citep[e.g.,][]{Wheeler15},
although more accurate method is to apply bolometric corrections %
\citep{Lyman14,Brown16}.} This treatment of the data from \cite{Sollerman02}
is not accurate enough because the multiplicative factors at different
phases should be different. In view of this fact we do not try to fit these
shifted data and upper limits in Figure \ref{fig:1998bw} and plot them here
just for eye guidance. The difference between the data from \cite{Patat01}
and \cite{Sollerman02} at similar times can be attributed to several
factors. First of all, the contribution from $UV$\ band is substantial and
cannot be neglected even at these advanced stages. Second, it was found that
the measured decay rates in $R$\ and $I$\ bands during the stage $300-500%
\unit{days}$ are different between \cite{Patat01} and \cite{Sollerman02}. 
\cite{Sollerman02} attributed this difference to different background
subtraction strategy. Third, \cite{Patat01} adopted a larger distance to SN
1998bw ($37.8\unit{Mpc}$) than \cite{Sollerman02} ($35\unit{Mpc}$). Finally,
because of the missing of measurements in $IR$\ band, these two groups of
authors assumed different contribution from $IR$\ band to the bolometric
luminosities.

Our modeling result is presented in Figures \ref{fig:1998bw} and \ref%
{fig:2002ap} with the best-fit parameters listed in Table \ref{tbl:para}. In
Figure \ref{fig:1998bw} we plot the data from \cite{Sollerman02} as dark
stars and two upper limits, after shifted upward by $0.6\unit{dex}$.

Figures \ref{fig:1998bw} and \ref{fig:2002ap} indicate that the early peaks
are caused by the magnetar injection, as already found by \cite{Wang16b}.
During the period from day 50 to day $\sim 300$ after explosion, the SN
light curve follows closely the $^{56}$Co exponential decay. After $\sim 300%
\unit{days}$ the light curve systematically deviates from this exponential
decay. What is surprising to us is that the late-time deviation from the
exponential decay can be curiously accounted for by the magnetar
contribution. This connection between early peak and late-time slow decay of
the light curve is unexpected in the two-component model \citep{Maeda03}.

In developing the magnetar model, to obtain an analytical result, the
temperature distribution within the ejecta is separated in space and time
coordinates \citep[see Appendix of][]{Wang16c}, $T\left( r,t\right)
^{4}\propto \psi \left( x\right) \phi \left( t\right) $, where $\psi \left(
x\right) $\ is a dimensionless function of mass coordinate $x$\ that
characterizes the radiation energy per unit volume. The heating rate is
similarly separated $\epsilon \left( r,t\right) =\xi \left( x\right) f\left(
t\right) $, where $\xi \left( x\right) $\ is another dimensionless function
that characterizes the volume emission after multiplied by the dimensionless
density $\eta \left( x\right) $. We assume that the volume emission of the
heating sources, including magnetar injection and $^{56}$Ni heating, is
proportional to the radiation energy per unit volume, i.e. $\xi \left(
x\right) \eta \left( x\right) \propto \psi \left( x\right) $ as suggested
originally by \cite{Arnett82} in developing an analytical model for ordinary
($^{56}$Ni only) type I SNe (assuming the quantity in Equation [13] of %
\citealt{Arnett82} a constant). This assumption is not strictly true, but it
is a good approximation which captures the main feature of thermalization of
the magnetar radiation and $^{56}$Ni decay photons.

When isolating the $^{56}$Ni contribution to the light curve, we actually
perform a subtraction between two light curves, the full light curve
(including magnetar and $^{56}$Ni) and the magnetar-only light curve. We do
not calculate the $^{56}$Ni light curve by turning off the magnetar
contribution because the magnetar could provide a substantial contribution
to the ejecta kinetic energy and the evolution of the ejecta expansion
velocity impacts the resulting light curve. Due to this fact, one is not
encouraged to directly compare the $^{56}$Ni contribution to a real $^{56}$%
Ni light curve.

There seems to be a little bump in the magnetar component around 150 days in
Figure \ref{fig:2002ap}. This bump occurs at the time when the SN is
transitioning to the full nebular phase. The rapid recession of the
photosphere cannot be handled accurately by the finite time step in the
numerical code. The real curve for the magnetar contribution should be very
smooth and one should therefore ignore this numerical artifact.

\section{Discussion}

\label{sec:dis}

There were attempts to reproduce the light curve of SN 1998bw by a
pure-magnetar model and pure-$^{56}$Ni model 
\citep[Figure 19
of][]{Inserra13}. The failure of these models calls for alternative models
for SNe Ic-BL, manifesting the investigation of magnetar plus $^{56}$Ni
model presented in this work.

The idea of combining a magnetar and $^{56}$Ni to reproduce the light curve
of SN 1998bw was previously discussed by \cite{Woosley10}. By considering
the luminosity of SN 1998bw at $1000\unit{days}$, \cite{Woosley10} estimated
that the magnetar would be born with a field strength in excess of $8\times
10^{15}\unit{G}$. It was estimated that a magnetar with such a strong field
will loss all of its rotational energy to explosion and none left to the
light curve. However, it is recently shown that such an estimate is only
partially correct \citep{Wang16b}. On the one hand, with a much strong
magnetic field, the magnetar is indeed prone to losing its energy to
explosion \citep{Wang16c}. On the other hand, however, a minor fraction of
the rotational energy will be thermalized, which is enough to power the
light curve \citep{Wang16b}.

In the magnetar (plus $^{56}$Ni) model, the intermediate-time ($50\lesssim
t\lesssim 300\unit{days}$) exponential decay, sensitive to the $^{56}$Ni
mass $M_{\mathrm{Ni}}$, in the light curve is caused by $^{56}$Co decay,
while in the two-component model this exponential decay is assumed to be
produced by the inner component. It is therefore desirable to compare the
value $M_{\mathrm{Ni}}$ in the magnetar model to the $^{56}$Ni mass of the
inner component in the two-component model. Comparison of Table \ref%
{tbl:para} with Table 2 of \cite{Maeda03} shows good agreement between $M_{%
\mathrm{Ni}}$ in the magnetar model and the inner component $^{56}$Ni mass $%
M\left( ^{56}\mathrm{Ni}\right) _{\mathrm{in}}$ in the two-component model.
We note that unusually strong nebular lines of [Fe {\scriptsize II}] are
seen in SN 1998bw, consistent with the production of more than the usual
amount of Fe in the explosion \citep{Mazzali01}.\footnote{%
The absence of [Fe {\scriptsize III}] in SN 1998bw \citep{Mazzali01},
however, is a concern because SNe Ia, which produce copious $^{56}$Ni, in
general show strong [Fe {\scriptsize II}] and [Fe {\scriptsize III}]
emission features. In addition, the wide bump in the region between $4000$
and $5500\unit{%
\text{\AA}%
}$ in the nebular spectra of SN 1998bw at days until about 337 post
explosion \citep{Patat01} precludes a large amount of $^{56}$Ni because
otherwise the line blanketing caused by iron group elements would
significantly suppress the radiation shortwards of $5000\unit{%
\text{\AA}%
}$ \citep{Dessart12,Dessart13}.} This is first of all in accord with the
magnetar model because the inferred $^{56}$Ni mass of SN 1998bw is indeed
larger than the $^{56}$Ni masses found in other SNe Ic-BL, e.g. SNe 1997ef,
2007ru, 2002ap, within the magnetar model. On the other hand, though, the
synthetic SN spectra suffer from large modeling uncertainties and a $^{56}$%
Ni mass of $0.1M_{\odot }$ is not inconsistent with observation. In the
two-component model, the inferred total $^{56}$Ni mass, $\sim 0.5M_{\odot }$%
, implies a progenitor core mass $\sim 3M_{\odot }$ %
\citep{Nakamura01a,Nakamura01b}, which is in excess of the maximum mass of a
neutron star, suggesting a collapsar (black hole) for GRB 980425. In the
magnetar model, however, the total $^{56}$Ni mass, $0.1M_{\odot }$, is
consistent with the formation of a magnetar. This also serves as a
self-consistent check for the magnetar model.

Except for the $^{56}$Ni mass, other parameters could be significantly
different between these two models. For SN 2002ap, the ejecta mass in the
magnetar model is in reasonable agreement with the total ejecta mass given
in the two-component model, while the ejecta mass of SN 1998bw in the
magnetar model is significantly smaller than that given by \cite{Maeda03}.
Table \ref{tbl:para} shows that, in the magnetar model, $M_{\mathrm{ej}%
}\left( \mathrm{SN\quad 1998bw}\right) \approx M_{\mathrm{ej}}\left( \mathrm{%
SN\quad 2002ap}\right) $. This result seems hard to understand. However,
when we compare the light curves of these two SNe \citep{Tomita06}, it is
immediately evident that both the early and late-time shape of the light
curve of SN 2002ap is strikingly similar to that of SN 1998bw.\footnote{%
Quantitative comparison shows some minor difference between these two SNe %
\citep{Cano13}.} In other words, SN 1998bw is just a brighter cousin to SN
2002ap. These two SNe have the same rise and decline rate around the peak
time. The fact that SN 1998bw has a slightly heavier ejecta mass is only
because, in the light curve modeling aspect, it expanded faster than SN
2002ap. Such a parameter difference for SN 1998bw between the two-component
model and the magnetar model may imply a difference for its progenitor. The
ejecta mass, $10M_{\odot }$, in the two-component model points to a massive
single star, while $M_{\mathrm{ej}}=2.6M_{\odot }$ in the magnetar model
favors a binary origin \citep{Fremling16}, although a $\sim 4M_{\odot }$
Wolf-Rayet star (here a remnant magnetar with typical mass $1.4M_{\odot }$
is assumed) evolved from a single $35M_{\odot }$ main sequence star is also
possible \citep{Woosley07,Woosley10}.

The initial kinetic energy in the magnetar model is $7.9\times 10^{50}\unit{%
erg}$ and $1.86\times 10^{51}\unit{erg}$ for SNe 2002ap and 1998bw,
respectively. These values are significantly smaller than that given by the
two-component model \citep{Maeda03} and in expectation from neutrino heating %
\citep{Janka16}.

The magnetic dipole field strength $B_{p}=\left( 13.7\pm 0.3\right) \times
10^{15}\unit{G}$ for SN 2002ap is similar to the value determined for SN
1997ef, while the value $B_{p}=\left( 1.66_{-0.14}^{+0.21}\right) \times
10^{15}\unit{G}$ for SN 1998bw is significantly smaller. The weaker magnetic
field for SN 1998bw is required because of its brighter luminosity %
\citep{Wang16b}. Usually a magnetar can have dipole field in the range $%
10^{14}-10^{15}\unit{G}$ \citep{Mereghetti08}. However, it is suggested that
a dipole field as strong as $\sim 10^{16}\unit{G}$ is possible in
theoretical aspects \citep{Wang16b}.

In this fitting a value $\kappa _{\gamma ,\mathrm{mag}}\gtrsim 2\unit{cm}^{2}%
\unit{g}^{-1}$ is favoured for SN 2002ap. But in practice it is found that $%
\kappa _{\gamma ,\mathrm{mag}}=2\unit{cm}^{2}\unit{g}^{-1}$ and $\kappa
_{\gamma ,\mathrm{mag}}=6\unit{cm}^{2}\unit{g}^{-1}$ result in essentially
the same light curve. This fact can also be appreciated by inspecting Figure %
\ref{fig:2002ap-corner}. From Table \ref{tbl:para} we see a large difference
of $\kappa _{\gamma ,\mathrm{mag}}$ between SNe 1998bw and 2002ap.
Inspection of Figure 8 of \cite{Kotera13} indicates that this could imply
average magnetar photon energies $\sim 10-100\unit{keV}$, in agreement with
observation \citep{Hester08}. Theoretically, magnetar radiation depends on
several parameters, e.g., magnetic field, rotation period, angle between
rotation axis and magnetic axis, and theorists are still struggling to
unequivocally predict its spectrum energy distribution 
\citep{Kennel84a,Kennel84b,Lyubarsky01,WangLJ13,
Kargaltsev15,Murase15,Wang15,Wang16a,LiuWang16}.

It can be seen that the two values of $\kappa _{\gamma ,\mathrm{Ni}}$ are
also different and larger than the fiducial value $\kappa _{\gamma ,\mathrm{%
Ni}}\approx 0.03\unit{cm}^{2}\unit{g}^{-1}$. We think of this as
macrophysical uncertainties, e.g., a clumpy density distribution. Radiation
hydrodynamic calculations show that the magnetar-driven ejecta pile up at
some radius \citep{Kasen10}, rather than homogeneously distributed, as
assumed in this work.

As stated above, the validity of the fitting program can also be appreciated
by comparing the determined value of explosion time $T_{\mathrm{start}}$
with the actual explosion time. For SN 1998bw whose explosion time is known,
the best-fit result gives $T_{\mathrm{start}}=-0.009_{-0.36}^{+0.32}\unit{%
days}$, in excellent agreement with observations. Such a consistency also
confirms the association of SN 1998bw with GRB 980425.

Inspection of Figures \ref{fig:1998bw} and \ref{fig:2002ap} shows that the
Monte Carlo program sensitively captures the shape changes in the light
curve. The decline rate of the light curve of SN 2002ap changes from $0.018%
\unit{mag}\unit{day}^{-1}$ between days 130 and 230 to $0.014\unit{mag}\unit{%
day}^{-1}$ between days 270 and 580 \citep{Tomita06}. There is a similar
change in the light curve of SN 1998bw between phase ranges 40-330 and
300-490 \citep[see Table 4 of][]{Patat01}. The flattening in the phase range
500-1200 of SN 1998bw (see Figure \ref{fig:1998bw}) is actually a
\textquotedblleft prediction" of the magnetar model because we do not fit
these data.

A number of effects were discussed that may contribute to this flattening,
including light echoes \citep{Cappellaro01,Andrews15,VanDyk15}, interaction
with circumstellar medium %
\citep{Chevalier82,Chevalier94,Ginzburg12,WangLiu16}, emission from
surviving binary star companion \citep{Kochanek09,Pan14}, radioactive
isotopes \citep{Woosley89,Sollerman02,Seitenzahl09}, clumping %
\citep{Maeda03,Tomita06}, positron escape \citep{Clocchiatti08,Leloudas09},
freeze-out of the steady state approximation \citep{Clayton92,Fransson93},
radiation transport for late SNe Ia \citep{Fransson15}, contribution from an
H {\scriptsize II} region or individual stars \citep{Patat01}, aspherical
geometry \citep{Maeda02,Maeda06}, gamma-ray trapping %
\citep{Clocchiatti97,Nakamura01a}, magnetar field decay \citep{Woosley10},
or contributions from collisions or a GRB afterglow \citep{Bloom99}.

Among all of the above possibilities, \cite{Sollerman02} considered a simple
but plausible case of $^{57}$Co and $^{44}$Ti radioactive contributions to
the flattening, although with a ratio of $^{57}$Ni/$^{56}$Ni three times
larger than the ratio observed in SN 1987A. Indeed, these radioactive
flattening resembles a magnetar powering at $t\approx 1200\unit{days}$
because of the progressive contributions from $^{56}$Co, $^{57}$Co and $%
^{44} $Ti at later phases \citep{Arnett89}.

In comparison to the above models, it was recently shown that in the
magnetar model $B_{p}$ and $P_{0}$ can be accurately determined by the early
peak for those SNe whose peak is caused by a spinning-down magnetar %
\citep{Wang16b}. The advantage of the magnetar model over the two-component
model is the fact that the magnetar parameters, $B_{p}$ and $P_{0}$,
determined by the early peak can naturally account for the late-time slow
decline in the light curves of SNe 1998bw and 2002ap. The magnetar initial
rotation periods $P_{0}\sim 20\unit{ms}$ for SNe 1998bw and 2002ap are
longer than that of SN 1997ef \citep{Wang16b}. This longer $P_{0}$ of the
magnetar is required for a relatively slow rise (compared to SNe 1997ef and
2007ru) of the peak in the light curve. This $P_{0}$ is also essential for
its contribution to the late-time observable excess relative to the $^{56}$%
Co exponential decay. In other words, the magnetar model predicts that the
rise rate in early-time light curve should be relatively slow if the SN has
a long lasting flattening. This intrinsic connection between the early peak
and late-time flattening is not expected in alternative models and therefore
provides a strong evidence in favor of the magnetar model.

It may be argued that the magnetar model presented here can give a better
account for these two SNe because there are more free parameters in this
model. This is not simply true at its first glance. There are 8 free
parameters in the magnetar model (see Table \ref{tbl:para}). In the
two-component model, the free parameters include $v_{\mathrm{in}}$, $M_{%
\mathrm{in}}$, $M\left( ^{56}\mathrm{Ni}\right) _{\mathrm{in}}$, $v_{\mathrm{%
out}}$, $M_{\mathrm{out}}$, $M\left( ^{56}\mathrm{Ni}\right) _{\mathrm{out}}$%
, $T_{\mathrm{start}}$. To account for the late flattening, provided that it
is attributed to $^{57}$Co and $^{44}$Ti, another two parameters, $M\left(
^{57}\mathrm{Co}\right) $ and $M\left( ^{57}\mathrm{Ti}\right) $, are
needed. It is 9 parameters in total.\footnote{%
Here we assume that in the two-component model, the opacities to gamma-ray
photons are taken the theoretical values and therefore are not free
parameters.}

SNe Ic-BL show some evidence of asphericity, especially those associated
with GRBs. In this case the early peak of the light curve may contain a
fraction of contribution by the fast-moving material in the ejecta. The peak
luminosity of an SN is sensitive to the dipole magnetic field $B_{p}$. The
contribution of the asphericity would increase $B_{p}$\ so that the magnetar
deposits less of its rotational energy into the light curve. Nevertheless
the observed polarization $0.5\%$\ of SN 1998bw at optical wavelengths can
be explained by an axial ratio less than $2:1$\ \citep{Hoflich95,Hoflich99}.
Such asphericity is actually comparable to the well-observed nearby SN 1987A %
\citep{Larsson16}. This indicates that the departure from spherical symmetry
of the optically emitting material is only moderate.

\section{Conclusions}

\label{sec:con}

The identification of the central engine of GRBs has long been a challenge
in high-energy astrophysics. Instead of investigating the prompt emission
and afterglows of GRBs, here we try to figure out what can be inferred by
studying the light curves of SNe Ic-BL. Stimulated by frequent hinds of
magnetar formation in GRBs and SNe Ic-BL, we present a magnetar (plus $^{56}$%
Ni) model for SNe Ic-BL and find evidence that SNe Ic-BL 1998bw and 2002ap
were powered by magnetars.

In more than one decade the picture of $^{56}$Ni heating to power the light
curves of SNe Ic-BL was developed, with two-component model the most
outstanding. Here we present evidence that magnetars could be a more natural
alternative. The two-component model does not account for the origin of the
huge kinetic energy of SNe Ic-BL. The magnetar model, on the contrary,
provides a complete solution for energetics, $^{56}\mathrm{Ni}$ synthesis,
and light curves. It is still to see if the spectra of SNe Ic-BL are
consistent with the magnetar model. We note that the nebular spectra of SN
1998bw at days until about 337 post explosion have a wide bump in the region
between 4000 and 5500$\unit{%
\text{\AA}%
}$ \citep{Patat01}. Such a feature is consistent with magnetar model.

To bring the magnetar model for SNe Ic-BL onto a more solid ground,
additional information is necessary. For example, the continuous injection
of magnetar energy at late time should affect the emission line width
evolution \citep{Chevalier92}. Future high accuracy multi-messenger
observations can also help identify the newborn magnetars in SNe Ic-BL %
\citep{Kashiyama16}.

We note in passing that another SN Ic-BL that was observed to late stage $%
t>300\unit{days}$ is SN 2003jd \citep{Valenti08}. However, because of the
sparse coverage of the observational data, we do not try to fit and analyze
it in this work. It is evident that SN 2003jd also displays a change in
decline rate between stages 50-100 and 300-400 \citep{Valenti08}, indicative
of magnetar formation. This may indicate that magnetar is a common remnant
in SNe Ic-BL.

\begin{acknowledgements}
We thank the anonymous referee for constructive comments.
This work is supported by the National
Basic Research Program (\textquotedblleft 973" Program) of China under Grant
No. 2014CB845800 and the National Natural Science Foundation of China (Grant
Nos. U1331202, 11533033, U1331101, 11573014, 11422325 and 11373022.).
D.X. acknowledges the support
of the One-Hundred-Talent Program from the National Astronomical
Observatories, Chinese Academy of Sciences.
\end{acknowledgements}

\appendix

\section{Parameter uncertainties}

\begin{figure*}[tbph]
\centering\includegraphics[width=0.88\textwidth,angle=0]{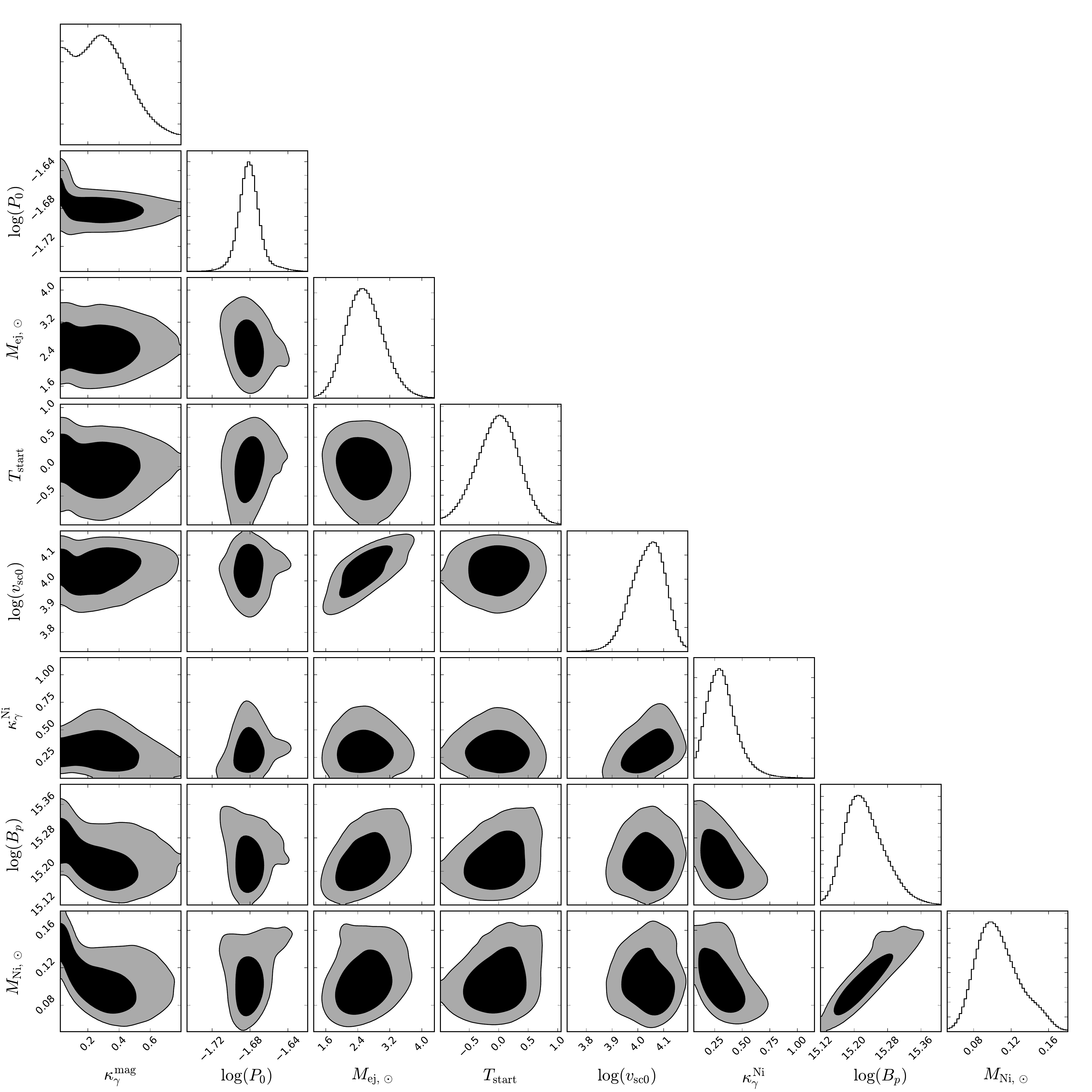}
\caption{Parameter corner in the modeling of SN 1998bw. The contours are $1%
\protect\sigma $ and $2\protect\sigma $ uncertainties, respectively.}
\label{fig:1998bw-corner}
\end{figure*}

\begin{figure*}[tbph]
\centering\includegraphics[width=0.88\textwidth,angle=0]{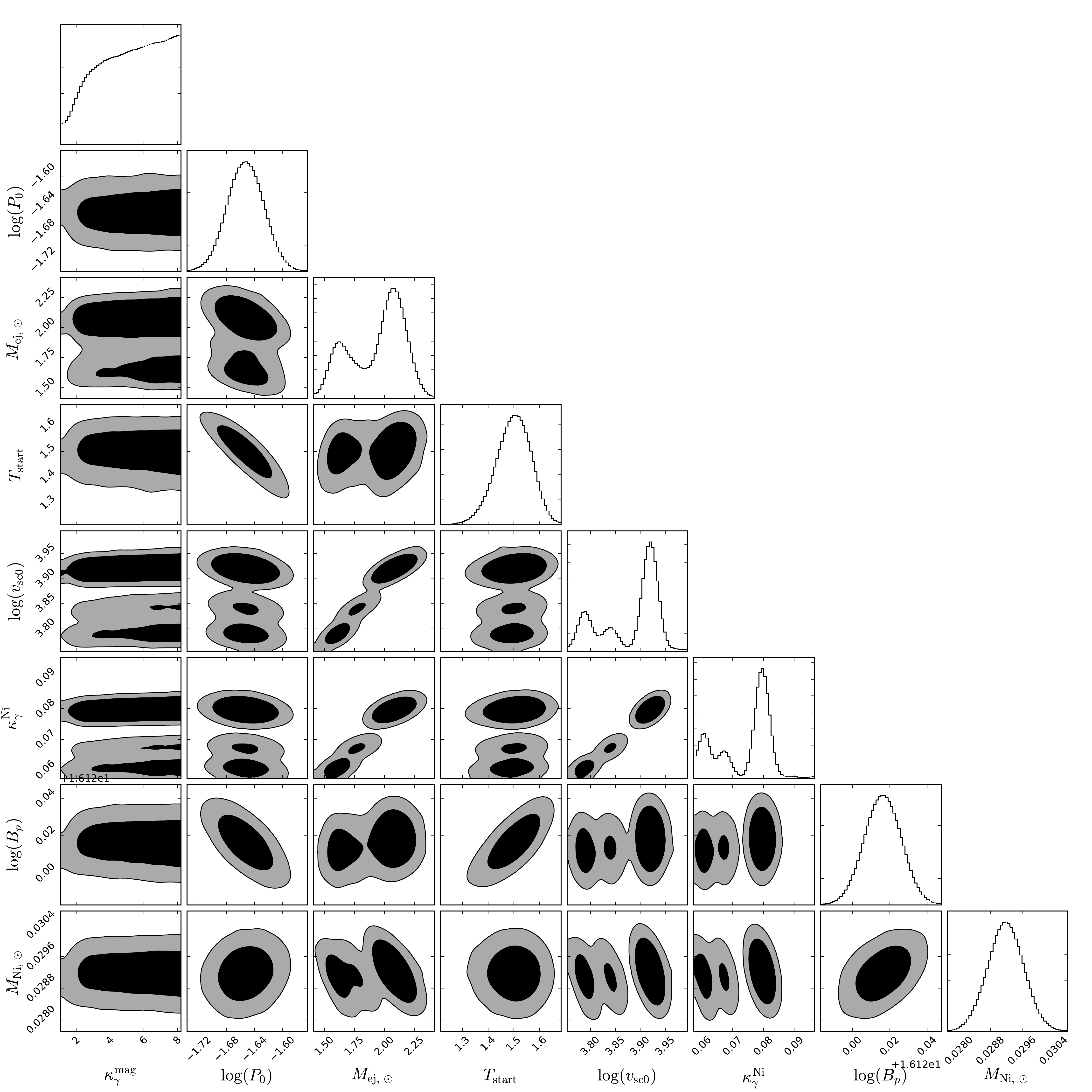}
\caption{Parameter corner in the modeling of SN 2002ap.}
\label{fig:2002ap-corner}
\end{figure*}

To help understand the parameter uncertainties and their degeneracy, we plot
the parameter corner graphs in Figures \ref{fig:1998bw-corner} and \ref%
{fig:2002ap-corner}. Figure \ref{fig:1998bw-corner} shows that $\kappa
_{\gamma ,\mathrm{mag}}\lesssim 0.5\unit{cm}^{2}\unit{g}^{-1}$ is favored
but it cannot be constrained tightly. This is because the observational
errors during the period 300-500 days are relatively large, as can be seen
from Figure \ref{fig:1998bw}. Figure \ref{fig:2002ap-corner} indicates some
degeneracy between $T_{\mathrm{start}}$ and $P_{0}$, $B_{p}$. This is easily
understood because $P_{0}$ and $B_{p}$ determine the rising rate of the
light curve. This figure also shows that $\kappa _{\gamma ,\mathrm{mag}}$
should not be less than $\sim 2\unit{cm}^{2}\unit{g}^{-1}$.

\end{document}